\documentclass[%
 reprint,
 amsmath,amssymb,
prb,
floatfix,
]{revtex4-2}
\usepackage[T1]{fontenc}
\usepackage{graphicx}
\usepackage{dcolumn}
\usepackage{bm}

\begin{document}

\preprint{APS/123-QED}

\title{Momentum considerations inside near-zero index materials}

\author{Micha\"el Lobet}
\altaffiliation{These authors contributed equally to this work.}
\affiliation {John A. Paulson School of Engineering and Applied Sciences, Harvard University, 9 Oxford Street, Cambridge, MA 02138, United States of America}
\affiliation{Department of Physics and Namur Institute of Structured Materials, University of Namur, Rue de Bruxelles 51, 5000 Namur, Belgium}
\email{michael.lobet@unamur.be}

\author{I\~nigo Liberal}
\altaffiliation{These authors contributed equally to this work.}
\affiliation{Electrical and Electronic Engineering Department, Universidad P\'{u}blica de Navarra,
Campus Arrosad\'ia, Pamplona, 31006 Spain}

\author{Larissa Vertchenko}
\altaffiliation{These authors contributed equally to this work.}
\affiliation{NanoPhoton - Center for Nanophotonics, Technical University of Denmark, Ørsteds Plads 345A, DK-2800 Kgs. Lyngby, Denmark}%

\author{Andrei V Lavrinenko}
\affiliation{NanoPhoton - Center for Nanophotonics, Technical University of Denmark, Ørsteds Plads 345A, DK-2800 Kgs. Lyngby, Denmark}%

\author{Nader Engheta}
\affiliation{Department of Electrical and Systems Engineering, University of Pennsylvania, Philadelphia, PA 19104, USA}

\author{Eric Mazur}
\affiliation {John A. Paulson School of Engineering and Applied Sciences, Harvard University, 9 Oxford Street, Cambridge, MA 02138, United States of America}
\email{mazur@seas.harvard.edu}

\date{\today}

\begin{abstract}

Near-zero-index (NZI) materials, i.e. materials having a phase refractive index close to zero, are known to enhance or inhibit light-matter interactions. Most theoretical derivations of fundamental radiative processes rely on energetic considerations and detailed balance equations, but not on momentum considerations. Because momentum exchange should also be incorporated into theoretical models, we investigate momentum inside the three categories of NZI materials, i.e. inside epsilon-and-mu near-zero (EMNZ), epsilon-near-zero (ENZ) and mu-near-zero (MNZ) materials. In the context of Abraham-Minkowski debate in dispersive materials, we show that Minkowski-canonical momentum of light is zero inside all categories of NZI materials while Abraham-kinetic momentum of light is zero in ENZ and MNZ materials but nonzero inside EMNZ materials. We theoretically demonstrate that momentum recoil, transfer momentum from the field to the atom and Doppler shift are inhibited in NZI materials.  Fundamental radiative processes inhibition is also explained due to those momentum considerations inside three-dimensional NZI materials. Lastly, absence of diffraction pattern in slits experiments is seen as a consequence of zero Minkowski momentum. Those findings are appealing for a better understanding of fundamental light-matter interactions at the nanoscale as well as for lasing applications. 

\end{abstract}

\maketitle


\section{Introduction}

In his seminal papers introducing fundamental radiative processes \cite{Einstein1916,Einstein1917}, Einstein noted that, while the description of the interaction between light and matter typically only take into account energy exchange, energy and momentum are directly connected to each other, and momentum exchange is equally important. A consequence of Einstein's theory of radiation is that the absorption/emission of a quantum of energy $\hbar \omega$ is accompanied by a momentum transfer $\hbar \omega/c=\hbar k$ between the field and the atom, with $\hbar$ the reduced Planck constant and $k$ being the wave vector. When an atom absorbs radiation, the momentum transfer is in the direction of propagation of the photon, while for emission the transfer is in the opposite direction, inducing a recoil of the atom.  
In a medium, the momentum of electromagnetic radiation (“electromagnetic momentum”) depends on the refractive index. However, there has been a long-standing debate concerning the dependence of the electromagnetic momentum on the refractive index depending on whether one uses the Minkowski \cite{Minkowski1910} or Abraham \cite{Abraham1909,Abraham1910} formulation of the electromagnetic momentum. The electromagnetic momentum density in the Abraham ($\bm{g}_A$) and Minkowski ($\bm{g}_M$) forms are

\begin{equation}
\bm{g}_A=\bm{E}\times\bm{H}/c^2
\label{Abraham density}
\end{equation}
and
\begin{equation}
\bm{g}_M=\bm{D}\times\bm{B}.
\label{Minkowski density}
\end{equation}
respectively \cite{Milonni2005,Garrison2014}. These formulations yield the following two expressions for the magnitude of the electromagnetic momentum in a dispersive medium:

\begin{equation}
p_A=\frac{\hbar\omega}{n_g(\omega) c},
\label{Abraham momentum}
\end{equation}
and 
\begin{equation}
p_M=n_\varphi(\omega)\frac{\hbar\omega}{c},
\label{Minkowski momentum 1}
\end{equation}
for the Abraham momentum $p_A$ and the Minkowski momentum $p_M$, respectively, and where $n_\varphi(\omega)=\sqrt{\varepsilon(\omega)\mu(\omega)}$ is the phase refractive index and $n_g=c (\,\frac{d\omega}{dk})\,^{-1}$ the group refractive index.

Note that in a non-dispersive medium, $n_\varphi=n_g=n$ and consequently $p_M=n\frac{\hbar\omega}{c}$ and $p_A=\frac{\hbar \omega} {n c}$. The difference between those two expressions for the electromagnetic momentum is at the heart of Abraham-Minkowski debate. Some experiments appear to support the Minkowski formulation \cite{Jones1954,Gibson1980,Campbell2005}, while others support the Abraham formulation \cite{Walker1975,Brevik1979,Pfeifer2007}.

A resolution of this long-lasting dilemma was recently proposed \cite{Barnett2010,BarnettLoudon2010} by attributing the difference between the Abraham and Minkowski momenta of light to the duality of light and matter \cite{Leonhardt2006}. For a particle, the classical (particle) momentum is given by the kinetic momentum, defined as $p_{kin}=p_A$. On the other hand, the canonical momentum, $p_C=h/\lambda$, embodies the wavelike nature of the particle. It was shown that in any light-matter interaction \cite{Barnett2010,BarnettLoudon2010}, the total momentum — a conserved quantity — is given by the sum of the kinetic momentum of the particle and the Abraham momentum of the light and is equal to the sum of the canonical momentum of the particle and the Minkowski momentum of the light:

\begin{equation}
 p_{kin}^{medium}+p_A=p_C^{medium}+p_M. 
\label{MomentumconsBarnett}
\end{equation}

One could therefore call the Abraham momentum the “kinetic momentum of the light” and the Minkowski momentum the “canonical momentum of the light” \cite{Barnett2010,BarnettLoudon2010}. In other words, the Abraham momentum comes into play when considering the particle nature of light and the Minkowski momentum when considering the wavelike nature of light \cite{Leonhardt2006}. In this paper we re-examine the difference between these two formulations in the extreme case that the index of refraction of the medium approaches zero.

\section{Momentum inside near-zero index materials}

In the past decade materials with near-zero refractive index have received a lot of attention because of their unusual optical properties, such as supercoupling \cite{Silveirinha2006,Silveirinha2007}, enhanced nonlinearities \cite{Reshef2019,Alam2016,Khurgin2021,Suresh2021} and fluorescence \cite{Fleury2013,Sokhoyan2013,Zheludev2020}, control of dipole-dipole interactions \cite{Sokhoyan2015,Liberal2016ScAd}, geometry-invariant resonant cavities \cite{Mahmoud2014}, photonic doping \cite{Liberal2017photonic} and propagation of the light power flow akin to ideal fluids \cite{LiberalPNASEMFluid}. The refractive index of a material is near zero when at least one of the two constitutive parameters of the refractive index — the relative electric permittivity $\varepsilon(\omega)$ or the relative magnetic permeability $\mu(\omega)$ — is close to zero \cite{Liberal2017,Vulis2019}. Near-zero index materials (NZI materials) fall into three categories: epsilon-near-zero (ENZ) materials where $\varepsilon$ approaches zero with nonzero $\mu$ \cite{Silveirinha2006,Edwards2008}; mu-near-zero (MNZ) materials with $\mu$ approaching zero with nonzero permittivity $\varepsilon$ \cite{Marcos2015}; or epsilon-and-mu-near-zero (EMNZ) media where both $\varepsilon$ and $\mu$ approach zero simultaneously \cite{Vulis2019, Ziolkowski2004,Mahmoud2014,Li2015,Briggs2013}. 

\subsection{Phase and group indices inside NZI materials}

At the zero-index frequency in a NZI materials, the phase index is zero, but it is important to note that the group index for an infinite, lossless material depends on the NZI materials category \cite{Liberal2017,Lobet2020}
\begin{equation}
  n_g(\omega=\omega_Z) = 
    \begin{cases}
      \infty & \text{ENZ \& MNZ materials}\\
      \omega_Z \ \partial_{\omega}n_\varphi \left(\omega_Z\right) & \text{EMNZ materials}
    \end{cases}       
    \label{Group index classes}
\end{equation}

Consequently, the group velocity $v_g$ is zero at the zero-index frequency in unbounded ENZ/MNZ materials \cite{Javani2016} but nonzero for EMNZ materials: $ \left (v_g(\omega=\omega_Z)=c/\omega_Z \partial_{\omega}n\left(\omega\right) \right)$ \cite{Lobet2020}. It should remain positive in low loss material, as imposed by causality.
Note that, despite having a near-zero group velocity, energy can be transmitted through a finite size ENZ(MNZ) sample \cite{Liberal2018Nanophot}. An exhaustive discussion on group and energy velocities inside infinite NZI materials sample is provided in Appendix \ref{AppendixVelocities}.

\subsection{Minkowski momentum inside NZI materials}

\begin{figure}[ht]
\centering\includegraphics[scale=0.45]{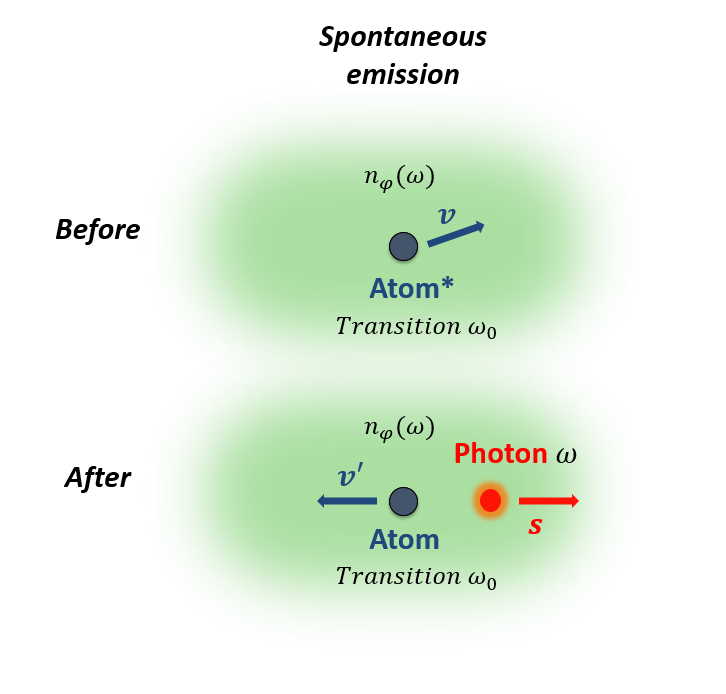}
\caption{Schematic of the spontaneous emission process inside a NZI materials material of refractive index $n_\varphi(\omega)$ (green background). An atom (blue) moves around at a velocity $v$ ($v'$) before (after) spontaneous emission process. A photon (red) is being emitted after the excited atom relaxed in its ground state, in direction $\bm{s}$.} 
\label{Dopplerdd}
\end{figure}

Because of the zero phase refractive index the Minkowski momentum — the canonical momentum of light — is zero for all NZI materials categories: $p_C=p_M=0$ see Eq. (\ref{Minkowski momentum 1}). Another way to show that the Minkowski momentum is zero involves applying the de Broglie relationship, $p_C=\frac{h}{\lambda}$, inside an NZI materials, which yields $p_C=0$,  because the effective wavelength $\lambda=\lambda_0/n_\varphi$ tends to infinity inside NZI materials, where $\lambda_0$ is the vacuum wavelength \cite{Liberal2017}. 
Another consequence is that no momentum is imparted by the photon to the material inside a NZI materials.  This point can be clarified using the example of the Doppler shift that occurs during spontaneous emission of radiation. Let us suppose an emitting atom of mass $m$, with a transition frequency $\omega_0$, an initial velocity $\bm{v}$ and a final velocity $\bm{v}'$ after emitting a photon of frequency $\omega$ (Fig. \ref{Dopplerdd}) \cite{Barnett2010,BarnettLoudon2010}.

In the non-relativistic approximation \cite{Fermi1932}, conservation of energy for the spontaneous emission process implies
\begin{equation}
\frac{mv^2}{2}+\hbar \omega_0 =\frac{mv'^2}{2}+\hbar \omega
\label{EnconsMink}
\end{equation}
while the conservation of linear momentum can be expressed as 

\begin{equation}
m\bm{v}=m\bm{v'}+\hbar \bm{k}
\label{pconsMink1}
\end{equation}

with $\bm{k}=\left[n_\varphi(\omega) \frac{\omega}{c}\right] \bm{s}$, $\bm{s}$ being an unit vector pointing in the direction of the emitted photon and $-\hbar \bm{k}$ is the recoil momentum of the atom. As is well known from classical physics, the frequency of the emitted light, as it appears to the moving atom, is increased due to the Doppler shift.
The Doppler shift formula can be deduced as \cite{Mansuripur2012,Mansuripur2012PRA}
\begin{equation}
\omega=\omega_0\left[1+\frac{n_\varphi(\omega)}{c}v cos\theta \right]
\label{Dopplershift}
\end{equation}
where $\theta$ is the angle between $\bm{v}$ and $\bm{s}$. 
 In general, $\hbar\bm{k}$ is not solely the momentum of the emitted photon, but corresponds to the total momentum transferred from the atom to both the emitted photon and the medium \cite{Milonni2005}.  The recoil of an emitter in a dispersive dielectric can be calculated either by using a macroscopic theory of spontaneous emission (considering the source atom as a two-level atom with a transition dipole $\bm{d}$) or by using field quantization in the dielectric \cite{Milonni2005}. Both approaches yield to the conclusion that the recoil momentum is the canonical momentum:
\begin{equation}
p_C=n_\varphi(\omega_0) \frac{\hbar\omega_0}{c}.
\label{Canonical momentum 2}
\end{equation}
Consequently, the recoil momentum vanishes inside NZI materials.
Moreover, the Doppler shift perceived by the atom also vanishes as the phase refractive index goes to zero (eq. \ref{Dopplershift}). This extinction of the Doppler shift can be understood as a continuous transition between inverse Doppler effect occurring in negative index materials \cite{Veselago1968,Kozyrev2005,Shi2018} and regular Doppler effect in positive index materials. Intuitively, inside NZI materials there is no phase difference, all parts of the material are tight to the same phase since the phase velocity is infinite. The compression or expansion of the wave fronts is not possible and the Doppler effect consequently vanishes. 
It should be noted here that local field corrections have no effect on the inhibition of the recoil momentum of the atom. Furthermore, a similar analysis can be done for deriving the recoil momentum in stimulated emission or in absorption processes \cite{Campbell2005} and will yield to the same conclusions in NZI materials. 
The absence of recoil momentum as a consequence of zero Minkowski momentum provides another way to understand inhibition of fundamental radiative processes inside three-dimensional NZI materials \cite{Lobet2020}. NZI materials forbids momentum exchange and the atom to recoil, in absorption and emissions processes. This can be seen as an environmental effect. This conclusion is totally consistent with our previous findings based solely on energy and detailed balance considerations \cite{Lobet2020}. Energy and momentum considerations are now treated on equal footing for the question of fundamental radiative processes as Einstein originally suggested \cite{Einstein1916,Einstein1917}. 
In summary, once wave aspects are dominating, marked in equations by the presence of the phase refractive index or the canonical momentum, related phenomena are inhibited inside NZI materials.

\subsection{Abraham momentum inside NZI materials}

\begin{figure}[h]
\centering\includegraphics[scale=0.42]{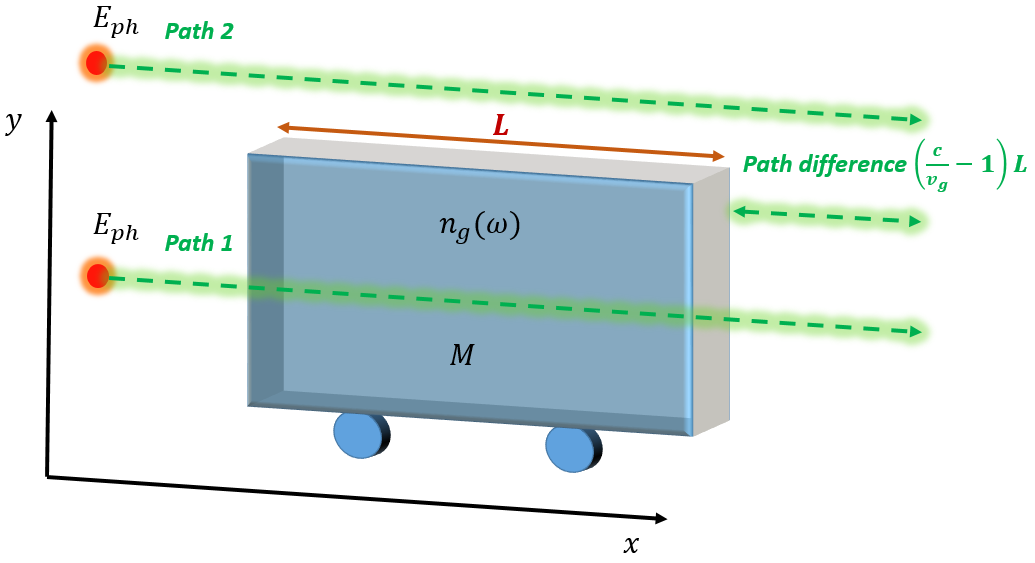}
\caption{Schematic of Balazs gedanken experiment inspired by \cite{Mansuripur2007OptExp,Mansuripur2011SPIE}. The photon can either take path 1 inside the slab of mass $M$, length $L$ and group index $n_g(\omega)$ or follow path 2 in free space. The slab is supposed to move freely, without friction in the $x$ direction.} 
\label{Balazs}
\end{figure} 
Consequences of near-zero refractive index on momentum considerations are different in particle-oriented experiments compared to wave-oriented experiments. Therefore, let us discuss one important particle-oriented experiment, the Balazs gedanken experiment \cite{Balazs1953}, applied to NZI materials. A detailed analysis of this gedanken experiment can be found in \cite{Mansuripur2011SPIE, Mansuripur2007OptExp,BarnettLoudon2010} and is partly reproduced in Appendix \ref{AppendixBalazs}. A photon propagates inside a transparent dielectric slab of length $L$, having a group refractive index $n_g(\omega)$ (Figure \ref{Balazs}). The slab can move without any friction along the $x$ axis and is supposed to be initially at rest ($v=0$). The photon propagates in the $x$ direction, enters the slab from the left facet and exits from the right facet. We suppose no losses due to absorption or scattering. The photon of energy $\hbar \omega $ propagates at velocity $c$ out of the slab (path 2), but propagates at the group velocity $v_g$ within the slab (path 1). 
By applying energy and momentum conservation laws, we can calculate the momentum gained by the slab $p_{slab}$ as well as the displacement of the slab $\Delta x$ due to the propagation of the photon following path 1. From there, we can deduce the momentum of the photon inside the slab, which reduces to the Abraham momentum (details in Appendix \ref{AppendixBalazs}) as given by eq. (\ref{Abraham momentum}). We remind here that the group index is the relevant one for the Abraham momentum and differs between ENZ/MNZ and EMNZ categories.

For ENZ/MNZ media, the group index is infinite (eq.(\ref{Group index classes})), and the group velocity is consequently zero \cite{Javani2016}. Therefore, the Abraham momentum is also zero (eq.(\ref{Abraham momentum})). For a sufficiently large lossless ENZ/MNZ slabs, the photon is completely reflected, and it bounces back at the material interface, communicating a forward momentum of $2\frac{\hbar \omega}{c}$ to an unbounded lossless ENZ/MNZ slab. However, inside an EMNZ medium, the group index is nonzero and equal to $\omega_Z \partial_{\omega}n_\varphi \left(\omega_Z\right)$ (eq.(\ref{Group index classes})). Therefore, propagation is allowed inside the slab that is displaced by a quantity $\Delta x_{EMNZ} =\left (\omega_Z  \partial_{\omega}n_{\varphi} \left(\omega_Z\right)-1 \right) L\frac{\hbar \omega }{Mc^2}$ (details in Appendix \ref{AppendixBalazs}). The slab acquires a momentum given by $p_{slab}=\left(1-\frac{1}{\omega_Z  \partial_{\omega}n_\varphi \left(\omega_Z\right)} \right)\frac{\hbar \omega }{c}$. 
Those considerations point a difference between EMNZ materials and photonic crystals. Experimental realization of EMNZ materials are photonic crystals showing a linear band dispersion around $\Gamma=0$, a crossing at the so-called Dirac point and a vanishing density of states at this point \cite{CTChang2011,Li2015,Vulis2019}. Even if spontaneous emission is forbidden inside such EMNZ photonic crystal \cite{Lobet2020}, propagation within EMNZ material is allowed. This is not the case for classical photonic crystals around their photonic bandgap \cite{Joannopoulos2008}. Consequently, photonic crystals with EMNZ properties allow both propagation of electromagnetic radiation and inhibition of spontaneous emission simultaneously, which are interesting for lasing platforms \cite{Bravo2012,Chua14,Faist2017} .

\section{Consequence of zero Minkowski momentum on diffraction}

Considerations on momentum inside NZI materials materials also have consequences on diffraction phenomena, e.g. on slit experiments inside dispersive media. Young double-slit experiments immersed inside a dielectric liquid can be found in literature, e.g. \cite{Ratcliff1972,Brevik1979,Brevik1981}. Recently, double-slit experiments were performed inside one of the NZI materials category, ENZ materials \cite{Ploss2017}. 
\begin{figure}[ht]
\centering\includegraphics[scale=0.65]{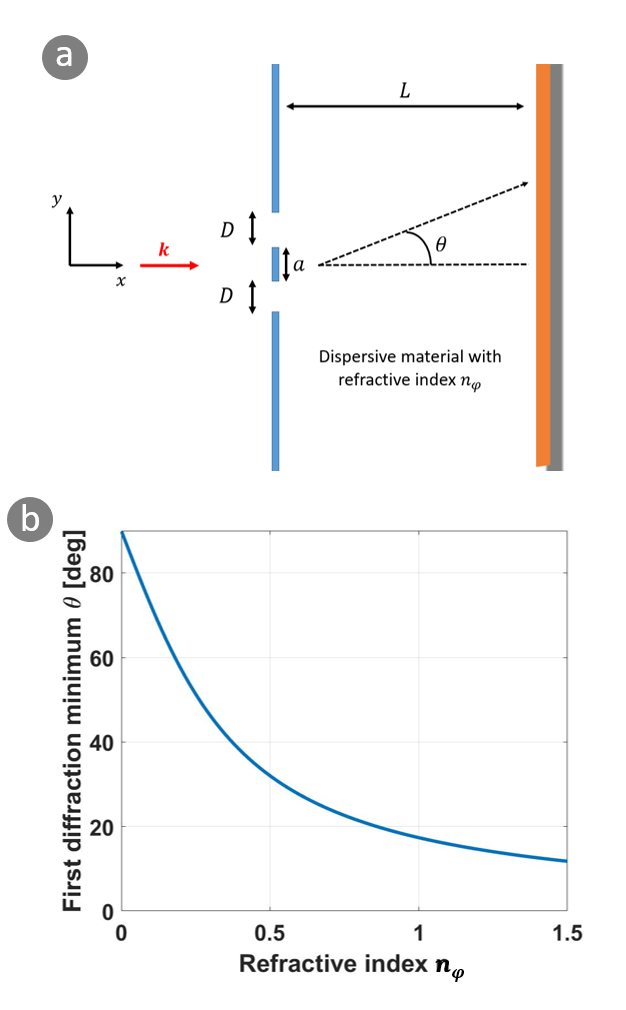}
\caption{(a) Young double-slit experiment within dispersive material. EM wave with $\bm{k}$ wavevector is sent on a double-slit with slit width $D$ at a distance $L$ from an observing screen. The distance between the slits is $a$.(b) First diffraction minimum as a function of refractive index. Wavelength is set to 500 nm for a separation width of $a=800$ nm.} 
\label{Young}
\end{figure}

Here, we consider a double-slit experiment, with a slit width $D$, separated by a distance $a$ (Figure \ref{Young}a). The distance between the double-slit and the observing screen is denoted as $L$ and the whole system, including the double-slit and the screen, are embedded inside a dispersive medium of refractive index $n_\varphi$. If $\theta$ denotes the angle between the forward $x$ direction and the direction of the first diffraction minimum, diffraction theory gives 
\begin{equation}
tan (\theta)=\frac{\lambda_0}{2a|n_\varphi|}
\label{Theta slit}
\end{equation}
if $L >> a$. In positive refractive index materials, the diffraction angle $\theta$ is consequently lowered by a factor $|n_\varphi|$ (Figure \ref{Young}b, for $n_\varphi >1$) while the corresponding canonical momentum $p_x$ is increased by the same factor. In NZI materials (Figure \ref{Young}b, for $n_\varphi<1$), the opposite situation occurs: the first diffraction minimum moves away from the $x$ axis as $n_\varphi$ decreases, while the canonical momentum tends towards zero. The localization in the momentum space imposes a delocalization in the position space, as a consequence of Heisenberg inequalities.
Moreover, the intensity distribution on the screen of this double-slit experiment follows
\begin{equation}
I(y)=\frac{I_0}{2} \frac{sin^2 \left( \frac{\pi D y}{\lambda L} \right)}{\left( \frac{\pi D y}{\lambda L} \right)} \left [ 1+cos \left( \frac{2\pi a y}{\lambda L} \right) \right]
\label{Intensity profile}
\end{equation}

\noindent with $I_0$ being the intensity of the incident wave. As the refractive index goes to zero, the effective wavelength $\lambda$ inside the medium goes to infinity, therefore the $cos$ term tends to one. Recalling that the $sinc$ function is equal to one at the zero-limit, the intensity on the screen appears to be constant, i.e. $I(y) \rightarrow I_0$. This calculation confirms that the first order diffraction minimum is removed to infinity and that diffraction effects are reduced in NZI materials. 
Same conclusions hold for single-slit experiments. It is interesting to note that regarding single-slit experiments, the suppression of diffraction pattern inside NZI materials is nothing but a consequence of Babinet's principle, i.e. diffraction pattern of a slit or of a rectangular object should be similar. Inside NZI materials, no scattering of objects can be identified rendering them invisible \cite{CTChang2011}. Cloaking corresponds to an infinite incertitude on the position of the invisible object, which can be reached using NZI materials as discussed above. 

\begin{figure}[ht]
\centering\includegraphics[scale=0.48]{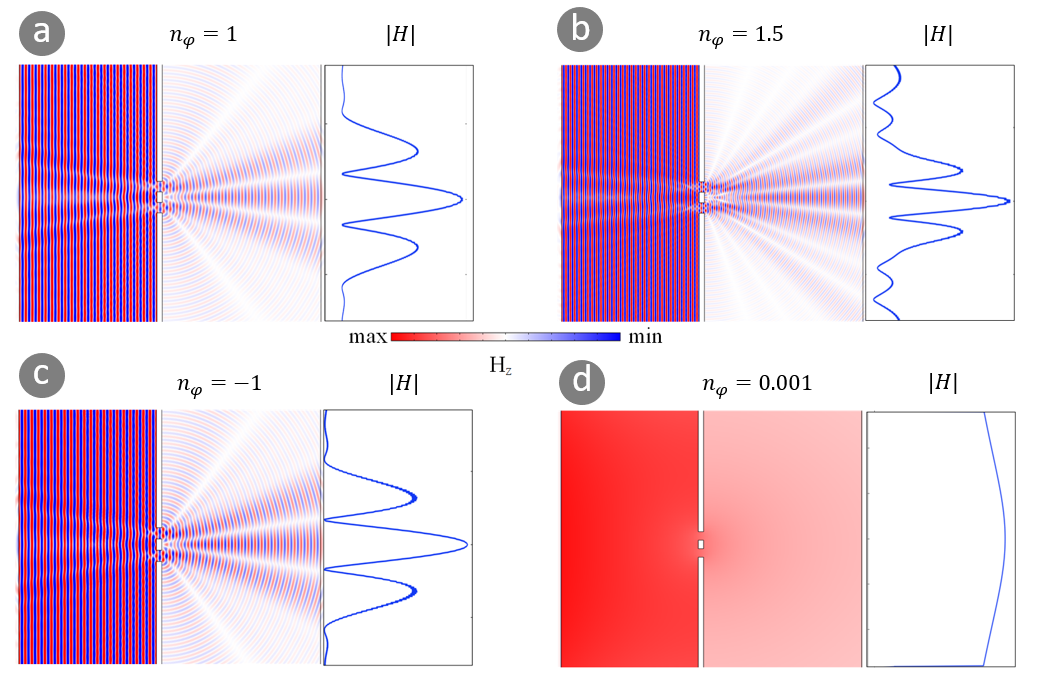}
\caption{Young double-slit experiment within different dispersive materials. Left: $|H_z|$ field maps, right: $|H|$ profile on the observing screen. (a) Air ($\varepsilon=\mu=1$,$n_\varphi=1$), (b) dielectric material such as glass ($\varepsilon=2.25, \mu=1$,$n_\varphi=1.5$), a negative refractive index material ($\varepsilon=-1, \mu=-1$,$n_\varphi=-1$) and  (d) an ENZ material ($\varepsilon=1\times 10^{-6}$, $\mu=1$ and $n=0.001$). Wavelength is set to 500 nm for a separation width of 800 nm.} 
\label{DiffractionH}
\end{figure}

The above theoretical considerations are verified by full-wave simulations \cite{Comsol} within four different materials: (a) Air ($\varepsilon=\mu=1$, $n_\varphi=1$), (b) dielectric material such as glass ($\varepsilon=2.25, \mu=1$, $n_\varphi=1.5$), a negative refractive index material ($\varepsilon=-1, \mu=-1$, $n_\varphi=-1$) and  (d) an ENZ material ($\varepsilon=1\times 10^{-6}$, $\mu=1$ and $n=0.001$). As we can observe on Figure \ref{DiffractionH}, the diffraction patterns (here, the $H$ field component) gets compressed inside dielectric material with refractive index higher than air (Figure \ref{DiffractionH}B), while no diffraction pattern appears within ENZ medium (Figure \ref{DiffractionH}D). The corresponding intensity profile on the screen and the direction of the first diffraction minimum are consistent with equations (\ref{Theta slit})-(\ref{Intensity profile}).

Finally, it is interesting to point that diffraction patterns are not influenced by the sign of the refractive index, but only by its absolute value (Figure \ref{DiffractionH}C). Consequently, the NZI materials scenario is an extreme case for diffraction theory as presented above.

\section{Conclusions}

Momentum considerations inside dispersive near-zero refractive index materials are theoretically worked out using the recent resolution of the Abraham-Minkowski debate \cite{Barnett2010,BarnettLoudon2010}. We evidenced that canonical-Minkowski momentum is identically zero inside NZI materials. This inhibits wave-related phenomena inside NZI materials. The Doppler shift perceived by the moving atom inside NZI materials is cancelled. No recoil momentum occurs inside such an unbounded lossless material. The dispersive material forbids the atom to recoil both in emission or absorption processes, leading to an absence of momentum exchange inside NZI materials. Fundamental radiative processes are inhibited inside three-dimensional NZI materials accordingly and this conclusion is consistent with the one derived using solely energetic considerations \cite{Lobet2020}. Energy and momentum are now treated on an equal footing regarding fundamental radiative processes inside NZI materials as Einstein suggested in seminal works \cite{Einstein1916,Einstein1917}. Absence of diffraction also appears as consequence of zero canonical momentum within NZI materials.
Nevertheless, for experiments where the corpuscular nature of light is probed, the Abraham momentum is linked to the group refractive index and therefore a distinction should be made according to the NZI materials category. Unbounded lossless ENZ/MNZ materials forbid direct propagation with zero kinetic-Abraham momentum, while bounded EMNZ allows direct propagation and nonzero kinetic-Abraham momentum. EMNZ-based photonic crystals can then be considered as specific materials allowing both light propagation but inhibiting spontaneous emission. This property is appealing for controlling fundamental radiative processes at the nanoscale as well for lasing perspectives.

\begin{acknowledgments}

The authors would like to thank Masud Mansuripur for constructive discussion on photon momentum inside dielectric. 
L.V acknowledges support from the Danish National Research Foundation through NanoPhoton - Center for Nanophotonics, grant number DNRF147. A. L. acknowledges the support from the Independent Research Fund Denmark, DFF Research Project 2 “PhotoHub” (8022-00387B), Villum Fonden.

\end{acknowledgments}

\appendix

\section{Energy density within dispersive and dissipative media and considerations on energy velocity} \label{AppendixVelocities}

We would like to establish a consistent formalism for the energy density within a dispersive and dissipative material. 
We will follow the Loudon approach \cite{Loudon2000}, well described by Nunes \cite{Nunes2011} and generalized to magnetic materials by Ruppin \cite{Ruppin2002}.

We analyse the energy density for a wave propagating through a dispersive and dissipative medium with a single resonant frequency. Dissipation can occur with both the E-field and the H-field \cite{Ruppin2002}. A simple model of a damped harmonic oscillator can be used to model dissipation, with $\omega_{0e}$ ($\omega_{0h})$ the resonance frequency of the electric (magnetic) dipole oscillators, $\gamma_e$ ($\gamma_m$) the damping rate and $\omega_{pe}$ ($\omega_{pm}$) characterizing the strength between the dipole oscillators and the electric (magnetic) field. 

The equation of motion of the electric polarization $\bm{P}$ in the presence of an oscillating electric field $\bm{E}$ can be written in the form

\begin{equation}
    \Ddot{\bm{P}}+\gamma_e \Dot{\bm{P}}+\omega_{0e}\bm{P}=\varepsilon_0 \omega_{pe}^2\bm{E}.
\end{equation}

Similarly, we can write the equation of motion of the magnetic polarization $\bm{M}$ in the presence of an oscillating magnetic field $\bm{H}$, in the form
\begin{equation}
    \Ddot{\bm{M}}+\gamma_m \Dot{\bm{M}}+\omega_{0m}\bm{M}=\mu_0 F\omega_{pm}^2\bm{H}.
\end{equation}
with $F$ a measure of the strength of the interaction between the oscillators and the magnetic field \cite{Ruppin2002}. 

For harmonic fields, keeping the $e^{-i\omega t}$ time dependence, the electric susceptibility $\chi_e$ (without dimensions) is defined by $\bm{P}=\varepsilon_0 \chi_e \bm{E}$.

The corresponding relative complex permittivity $\tilde{\varepsilon}_r(\omega)$ (dimensionless) is defined as

\begin{align}
    \tilde{\varepsilon}_r(\omega)&=\varepsilon_r(\omega)'+i\varepsilon_r(\omega)''=1+\chi_e= 1-\frac{\omega_{pe}^2}{\omega^2-\omega_{0e}^2+i\gamma_e\omega} \\ &= 1-\frac{w_{pe}^2(\omega^2-\omega_{0e}^2)}{(\omega^2-\omega_r^2)^2+\gamma_e^2\omega^2} - i \frac{\omega_{pe}^2\gamma_e\omega}{(\omega^2-\omega_{0e}^2)^2+\gamma_e^2\omega^2} 
    \label{ComplexPermittivity}
\end{align}
and is related to the permittivity $\tilde{\varepsilon}(\omega)$ [$F/m=\frac{As}{Vm}$] by $\tilde{\varepsilon}(\omega)=\varepsilon_0\tilde{\varepsilon}_r(\omega)$.

Similarly, the magnetic susceptibility $\chi_m$ is defined by $\bm{M}=\chi_m \bm{H}$ and the corresponding relative complex permeability writes
 \begin{align}
   \tilde{\mu}_r(\omega)&=\mu_r(\omega)'+i\mu_r(\omega)''=1+\chi_m=1-\frac{F\omega_{pm}^2}{\omega^2-\omega_{0h}^2+i\gamma_m\omega} \\ &= 1-\frac{w_{pm}^2(\omega^2-\omega_{0m}^2)}{(\omega^2-\omega_{0m}^2)^2+\gamma_m^2\omega^2} - i \frac{\omega_{pm}^2\gamma_m\omega}{(\omega^2-\omega_{0m}^2)^2+\gamma_m^2\omega^2}
   \label{ComplexPermeability}
 \end{align}
 and is related to the permeability $\tilde{\mu}(\omega)$ [$H/m]$] by $\tilde{\mu}(\omega)=\mu_0\tilde{\mu}_r(\omega)$.
  Consequently, we have the complex refractive index $\tilde{n}=\sqrt{\tilde{\varepsilon_r} \tilde{\mu_r}}=n'+in''$, the complex wave vector $\tilde{\mathbf{k}}=\mathbf{k_0} \tilde{n}$ and the group velocity $v_g=d\omega/dk$. We should note that near an absorptive resonance, the group velocity might be greater than the speed of light (anomalous dispersion) calling for a definition of energy velocity: $v_E=\frac{\mathbf{S} }{\overline{w}}$ with $\overline{w}=\overline{w_e}+\overline{w_m}$ the time-averaged total energy density. Loudon recasted the energy velocity within a lossy material as the ratio of the Poynting vector (time rate of energy flow normal to a unit area) and the total energy density.
The above definition leads to the most general form of the total energy density in a medium having both permittivity and permeability dispersive and absorptive. The total energy density $w$ comprises four terms for the stored energy $w_s$ and the dissipative energy $w_d$ for both the E and H-fields. The time-average of the total energy density $\overline{w}$ writes \cite{Ruppin2002,Nunes2011}: 

\begin{equation}
\boxed{\overline{w}  = \frac{\varepsilon_0}{4} \left( \varepsilon'_r + \frac{2\omega\varepsilon_r''}{\gamma_e} \right) |\mathbf{E}|^2 +\frac{\mu_0}{4} \left(\mu_r' + \frac{2\omega\mu_r''}{\gamma_m} \right) |\mathbf{H}|^2}.
\label{TotalEnDens}
\end{equation}

This is the most general form of the energy density in a medium in which both the permittivity and permeability are dispersive and absorptive. 
The total energy density $w$ is the sum of the electric and magnetic energy densities and each comprise two terms: a \textit{stored energy} term $w_s$ and a \textit{dissipative energy} term $w_d$ respectively coming from the real and imaginary parts of the permittivity/permeability. It was interestingly described in such a way by Webb in 2012 \cite{Webb12}.  We have 

\begin{equation}
    w=w_s+w_d=w_{se}+w_{sm}+w_{de}+w_{dm}.
\end{equation}

We can take the free-electron gas limit, $\omega_{0e} \rightarrow 0$ where the restoring force on the oscillating electrons goes to zero \cite{Nunes2011}. It is nothing but the Drude model of metals. In this limit, we have $\omega \tau \gg 1$, with $\tau$ the scattering time, we have 
\begin{equation}
    \varepsilon'_r(\omega)\rightarrow 1-\frac{\omega_{pe}^2}{\omega^2};
\end{equation}

\begin{equation}
    \varepsilon''_r(\omega)\rightarrow \frac{\gamma_e\omega_{pe}^2}{\omega^3}.
\end{equation}

From these last two expressions, we have

\begin{equation}
\frac{d\varepsilon_r'}{d\omega}=\frac{2\omega_{pe}^2}{\omega^3}
   \quad\mathrm{and}\quad 
\omega\frac{d\varepsilon_r'}{d\omega}=\frac{2\varepsilon_r''\omega}{\gamma_e}.
\label{depsdw}
\end{equation}

Within this limit, Loudon's approach is equivalent to Brillouin's approach \cite{Nunes2011}. 
 
Another interesting limit is the one of low losses. It happens when the damping term, controlled by $\gamma_e$ ( respectively $\gamma_m$), goes to zero. According to the equation \ref{ComplexPermittivity} (eq. \ref{ComplexPermeability}), the imaginary parts go to zero ($\varepsilon_r''\rightarrow 0$ and $\mu_r'' \rightarrow 0$) as expected. Moreover, we can check that for the permittivity \footnote{Equivalent relations can be derived for permeability.}, we have

\begin{equation}
    \frac{2\varepsilon_r''\omega}{\gamma_e}\rightarrow \frac{2\omega_{pe}\omega^2}{(\omega^2-\omega_{0e}^2)^2}
\end{equation}
and 
\begin{equation}
    \lim_{\gamma_e\to 0}\frac{d\varepsilon_r'}{d\omega}= \frac{2 \omega_{pe}\omega}{(\omega^2-\omega_{0e}^2)^2}.
\end{equation}

Consequently, the stored energy density is the limit of eq. \ref{TotalEnDens} for small losses ($\gamma \rightarrow 0$ and yields 
 
 \begin{align} 
\lim_{\gamma\to 0} \overline{w}  &= \overline{w_S} \\ 
 &=  \frac{\varepsilon_0}{4} (\ \varepsilon_r' + \omega \frac{d\varepsilon_r'}{d\omega} )\ |\mathbf{E}|^2 +\frac{\mu_0}{4} (\mu_r' + \omega \frac{d\mu_r'}{d\omega} )\ |\mathbf{H}|^2.
\label{Stored energy}
 \end{align}
 The frequency derivatives taking dispersion into account, even if the material does not absorbs intensively. Nunes adds that the dissipated power density \cite{Nunes2011}
 \begin{equation}
     \frac{\gamma_e\omega^2}{\varepsilon_0\omega_{pe}^2}=\varepsilon_0\varepsilon''_r\omega |\mathbf{E}|^2
 \end{equation}
 also goes to zero in this limit.
 
 If the dissipation is not negligibly small, the term including the dissipative energy density, averaged over an optical cycle is
 \begin{equation}
 \overline{w_d} = \frac{\varepsilon_0}{2} \frac{\omega\varepsilon_r''}{\gamma_e} |\mathbf{E}|^2 +\frac{\mu_0}{2} \frac{\omega\mu_r''}{\gamma_m} |\mathbf{H}|^2
 \label{Dissipative energy}
\end{equation}
 
 It is worthwhile mentioning that this dissipative energy density does not go to zero as $\gamma$ does (it is logical since we did not calculate this limit). We can define an average dissipative energy density $\bar{w_d}$ as the product of the cycle-averaged dissipative power density ($\varepsilon_0\varepsilon''\omega |E|^2 + \mu_0 \mu''\omega |H|^2$) and the period of a cycle:
 
 \begin{equation}
 \bar{w_d}  = \frac{1}{2} \varepsilon_0 \varepsilon'' |\mathbf{E}|^2 +\frac{1}{2} \mu_0 \mu'' |\mathbf{H}|^2
 \label{Dissipative energy average}
\end{equation}
where it clearly vanishes when $\varepsilon''$ and $\mu''$ goes to zero. 

Therefore, inside dispersive and dissipative NZI materials materials, we can control the stored energy density via the real parts of $\varepsilon$ and $\mu$ while the dissipative energy density is controlled by the imaginary parts. The NZI materials limits lead to an absence of stored energy, which is fully consistent with the absence of momentum transfer inside NZI materials. Only dissipative energy survives.
If losses are small, one retrieves our previous case of zero energy density in the NZI materials limit \cite{Lobet2020}

Ruppin gives an energy velocity equals to 

\begin{equation}
 v_E= \frac{2c Re \big( \sqrt{\frac{\varepsilon}{\mu}}\big )} {(\varepsilon' + \omega \frac{d\varepsilon'}{d\omega} )\  + (\mu' + \omega \frac{d\mu'}{d\omega}) }.
 \label{Energy velocity}
\end{equation}

In the EMNZ limit (dispersive and dissipative), we get

 \begin{equation}
 v_E= \frac{2c Re \big( \sqrt{\frac{\varepsilon}{\mu}}\big )} {\omega \big( \frac{d\varepsilon}{d\omega} + \frac{d\mu}{d\omega} \big | \frac{\varepsilon}{\mu}\big | \big) }
 \label{Energy velocity EMNZ}
\end{equation}

which leads, in case of a low loss limit to

 \begin{equation}
 v_E= \frac{c}{\omega \frac{dn}{d\omega}}=v_g.
 \label{Energy velocity EMNZ2}
\end{equation}

\section{Balazs gedanken experiment} \label{AppendixBalazs}

First, let us discuss Balazs gedanken experiment [Balazs1953], well described in [MansuripurSPIE2011,BarnettLoudon2010]. We can have the following reasonging either with a short light pulse or using a photon [MansuripurSPIE2011]. We choose here the photon version.

Let us consider a transparent dielectric slab of length $L$, having a group refractive index $n_g(\omega)$ (Figure \ref{Balazs}). The slab can move without any friction along the $x$ axis and is supposed to be initially at rest ($v=0$). The photon propagates in the $x$ direction, enter the slab from the left facet and exits from the right. We suppose no losses due to absorption or scattering. It possess an energy $\hbar \omega $ and propagates at velocity $c$ outside the slab. If one applies energy conservation before the photon enters the slab, the system (i.e. the slab+the photon) has a total energy

\begin{equation}
E_{tot}=E_{phot,vac}+E_{slab,vac}=\hbar\omega+Mc^2
\label{Energycons}
\end{equation}
where $M$ stands for the mass of the slab. The subscript $vac$ indicates that the photon is within vacuum at that time.
The aim of the present consideration is to find the electromagnetic momentum of the photon inside the slab. 
 Upon each encounter with the facets, the photon can either get reflected at the interface or go through the slab. Consequently, there will be an infinite number of possible outcomes for the present gedanken experiment. As Mansuripur did [MansuripurSPIE2011], we will examine only three significant cases:
\begin{itemize}
  \item The photon bounces back from the first facet;
  \item The photon goes through both facets successively, without any reflection at all;
  \item The photon enters the slab, bounds two times are the right and left facets successively then exists at the right facet. 
\end{itemize}

The simplest possibility is the first situation where the photon is directly reflected backwards at the slab interface. Its vacuum momentum $\frac{\hbar\omega}{c}$ reverses its direction and following momentum conservation, the slab acquires a forward momentum of $2\frac{\hbar\omega}{c}$. Let us recall here that momentum conservation implies that, for any system not subjected to external forces, the momentum of the system (here slab+photon) will remain constant. It means that the center of mass, or more precisely the centre of mass-energy, moves with a constant velocity [Einstein1906]. In this first case, the center of mass-energy of the system continues to move forward at the same rate as if the photon was travelling outside the slab (path 2). It is impossible to learn anything about the EM momentum inside the medium since the photon never entered the slab.

The second case is much more appealing since the photon propagates inside the slab over a distance $L$. Once it enters the medium, its speed slows down to $c/n_g(\omega)$ since, if there are no losses, the energy velocity is the group velocity $v_g(\omega)$ (See Appendix \ref{AppendixVelocities}). It consequently takes the photon a time $\Delta t=n_g(\omega)L/c$ to travel through the medium. Therefore, when the photon emerges from the slab following path 1, it will be delayed due to this reduced velocity compared to a photon travelling on path 2, outside the slab. In other words, a photon following path 2 is ahead a distance $(\,n_g(\omega)-1)\,L=(\,\frac{c}{n_g(\omega)}-1)\,L$ compared to the photon emerging from the slab on path 1.
The deviation from uniform motion can consequently only be made up if the block itself is displaced in the direction of propagation of the photon by an amount $\Delta x$ while the photon is in the medium.  This displacement can be calculated as follows. The delay has caused a leftward shift of the product of mass by displacement to the photon:

\begin{equation}
\frac{E_{phot,vac}}{c^2}\left(n_g(\omega)-1)\right)L=\frac{\hbar\omega}{c^2}\left(n_g(\omega)-1)\right)L
\label{Massdispl}
\end{equation}
where $\frac{\hbar\omega}{c^2}$ is the corresponding mass of the photon. This shift must be compensated by a rightward shift of the slab itself: 
\begin{equation}
\Delta x M.
\label{Massdisplslab}
\end{equation}
Equating both leftward and rightward mass times displacements yields
\begin{equation}
\Delta x =\left(n_g(\omega)-1)\right) \frac{\hbar\omega L}{Mc^2}.
\label{Massdisplslab2}
\end{equation}
We can clearly observe that this displacement depends linearly on the thickness of the slab, the ratio of the photon and material energies and the group refractive index. If the medium would have been made of vacuum, $\Delta x=0$. Consequently, the slab acquired a momentum from the photon in order to move the distance $\Delta x$:

\begin{equation}
p_{slab}=Mv=M\frac{\Delta x}{\Delta t}=\left( 1-\frac{1}{n_g(\omega)} \right) \frac{\hbar\omega}{c}.
\label{Massdisplslab3}
\end{equation}
Now, by simply applying momentum conservation to the system, we deduce the momentum of the photon inside the slab $p_{phot,slab}$:
\begin{equation}
p_{phot,vac}+p_{slab,vac}=p_{phot,slab}+p_{slab,slab}
\label{Momentumbalazs}
\end{equation}
and considering that the slab is initially at rest, it leads to
\begin{equation}
p_{phot,slab}=\frac{\hbar\omega}{cn_g(\omega)}=p_A.
\label{Momentumbalazs_final}
\end{equation}
We see that the momentum of the photon inside the slab is nothing but Abraham momentum.

The third case can be treated in a similar way, reaching the same conclusion for the EM momentum. If the photon enters the slab from the left, bounces back two times at the facets then exists on the right facet, it will have spent a total time of $\Delta t'=3 n_g(\omega)L/c$ seconds inside the slab. Its delay will be $\left(3n_g(\omega)-1\right)L\frac{\hbar \omega }{c^2}$ compared to free-space photon on path number 2. Equating the leftward and rightward mass times displacement of both the photon and the slab yields $p_{phot,slab}=\frac{\hbar\omega}{cn_g(\omega)}$ in any cases for the EM momentum inside the dielectric.


\bibliography{Library}

\end{document}